\documentclass[aps,prl,twocolumn,showpacs]{revtex4-1}
\usepackage{amsfonts}
\usepackage{amsmath}
\usepackage{amsthm}
\usepackage{graphics}
\usepackage{graphicx}
\usepackage{subfigure}
\usepackage{amssymb}
\usepackage{graphics}
\usepackage{graphicx}
\usepackage{epstopdf}
\usepackage{color}
\usepackage{subfigure}
\usepackage{pgfplots,tikz}
\usepackage{url}
\usepackage{float}
\usepackage{soul}

\definecolor{pink}{rgb}{1,1,0} 
\definecolor{red}{rgb}{1,0,0}
\definecolor{blue}{rgb}{0,0,1}
\definecolor{green}{rgb}{0,1,0}
\definecolor{yellow}{rgb}{1,1,0}
\definecolor{orange}{rgb}{1,0.5,0}
\definecolor{white}{rgb}{1,1,1}

\begin{document}

\title{Is turbulence universal ? A Kolmogorov spectrum for strongly vibrating plates.}

\author{Gustavo D\"uring}
\affiliation{Facultad de F\'isica, PUC, Chile.}
\author{ Christophe Josserand}
\affiliation{LadHyX, CNRS \& Ecole Polytechnique, UMR 7646, 91128, Palaiseau, France.}
\author{ Giorgio Krstulovic}
\affiliation{Universit\'e C\^ote d'Azur, Observatoire de la C\^ote d'Azur, CNRS, Laboratoire Lagrange, Bd de l'Observatoire, CS 34229, 06304 Nice cedex 4, France.}
\author{Sergio Rica}
\affiliation{Facultad de Ingenier\'ia y Ciencias, Universidad Adolfo Ib\'a\~nez, Santiago, Chile.}

\begin{abstract}
In fluid turbulence, energy is transferred from a scale to another by an energy cascade that depends only on the energy dissipation rate. It leads by dimensional arguments to the Kolmogorov 1941 (K41) spectrum.  Remarkably  the normal modes of vibrations in elastic plates manifests an energy cascade  with the same K41 spectrum in the fully non-linear regime.
 Moreover, the elastic deformations present  large ``eddies'' together with a myriad of small ``crumpling eddies'', such that folds, developable cones, and more complex stretching structures, in close analogy with spots, swirls, vortices and other structures in hydrodynamic turbulence. 
We characterize the energy cascade, the validity of the constant energy dissipation rate over the scales and the role of intermittency via the correlation functions.
 \end{abstract}

\maketitle

Turbulence has remained a central problem in fluid dynamics since the early experiments of Osborne Reynolds \cite{Reynolds}. Perhaps the most salient feature is the setting of an energy cascade that redistributes the energy among different Fourier modes of the velocity fluctuations that are independent of the dissipation caused by viscosity at small scales. This energy cascade depends only on the energy dissipation rate per unit mass, $P$ and the wave number $k$ leading to the Kolmogorov K41 \cite{Kolmogorov} spectrum: $E_k \sim P^{2/3} k^{-5/3}$. Despite a century of effort, turbulence still remains nowadays a major challenge from the experimental, theoretical and numerical points of view, and the very essence of the phenomena has not yet been revealed satisfactorily \cite{JimenezScience,NatureComment,Ruelle}. For instance, if the K41 law can be deduced using dimensional arguments, the statistical description of the turbulence fails to describe correctly the general correlation functions of the velocity fields. In fact, up to date, there is only a single analytical result that maybe derived  from the original Navier-Stokes (NS) equations for incompressible fluids, the so-called von K\'arm\'an-Howarth relation \cite{vKH}, that links the second and the third order correlations function of some component of the fluid velocity with the energy dissipation rate. Higher moments of these quantities differ from mean-field based predictions, and experimental as well as numerical data reveal extreme events, exhibiting heavy tailed distributions, that are usually associated with {\it intermittency}. Moreover, one can argue that the difficulty of describing turbulence may come from the fundamental issues raised by the fluid mechanics equations, witnessed in particular by the Clay millennium problem on the regularity of the NS equation in three space dimensions. Similarly, fluid turbulence is a hard problem because it is in the same footing that the Euler equation for perfect fluids which is {\it ab-initio} a nonlinear partial differential equation without any small parameter that may justify an asymptotic scheme nor a rational closure for the statistical correlations \cite{FrischBook}. Since fluid turbulence is studied in the framework of the  NS equations, it is important to question whether such cascade dynamics transporting the  energy from the large injection scales to the small dissipative ones can be observed in different contexts and whether the turbulence features described in fluids are universal or dependent on the dynamical equations. In particular, while the quadratic nonlinear term of the NS equation is responsible of the mixing between scales at the heart of the cascade process, nothing in the theoretical description of turbulence takes into account the details of the nonlinearity itself. 
In fact, turbulence has been observed in different contexts such as Magneto-Hydrodynamics where the magnetic field is coupled with the flow
or viscoelastic turbulence in polymer solutions at low Reynolds numbers 
for instance.  In these cases even though the nonlinearity differs from the one of NS equations and different 
spectra can be measured, they have similar properties than the original Kolmogorov phenomenology.

On the other hand, a different class of turbulence exists when a linear term is present in the dynamics, the so-called wave turbulence. Wave propagation dominates at linear order and the waves interact through the nonlinear terms. The dynamics is described by the weak or wave turbulence theory (WTT), developed originally in the sixties~\cite{hasselmann,benney,ZakhBook,newell}. WTT is deduced in the limit of weak wave amplitudes, so that the nonlinear terms can be treated perturbatively, by contrast with fluid turbulence where the nonlinearity is present at first order in the dynamics~\cite{ZakhBook,NewellRumpf,NazBook}. Based on the long-time statistics of randomly fluctuating interacting waves, the WTT deduces a kinetic equation for the distribution of spectral densities. Beside thermodynamical equilibrium solution, the key-feature of this equation is that non-equilibrium turbulent stationary solutions also arise, called in general Kolmogorov-Zakharov spectra (KZ). 
Indeed, as shown by Zakharov~\cite{ZakhBook}, they
describe a constant flux transfer (or cascade) of conserved quantities ({\it e.g.} energy) between large and small length scales. Examples of wave turbulence go much beyond the cases of surface gravity or capillary waves and concerns systems as diverse as plasma waves~\cite{ZakhBook}, nonlinear optics~\cite{Dyachenko-92}, vibrating elastic plates \cite{during} and  gravitational waves in general relativity~\cite{GravWavesNazarenkoGaltier2017},  among others. Although the analogy with turbulent flows is natural since it describes a flux of the
energy between scales, the links within the physical mechanisms is less straightforward: in particular, the WTT is based on the resonant interactions between the waves prescribed by the nonlinearity so that the power of the nonlinearity is crucial in the process. For instance if the pertinent nonlinearity is quadratic, the dynamics is determined by three waves interactions such that both the wave-numbers and the frequencies of the three waves form resonant triads. This mechanism implies a specific dependence of the KZ spectrum with the energy flux $P$ as $P^{1/N}$ where $N$ is the power of the pertinent nonlinearity of the dynamics ($N=2$ for quadratic interactions, $3$ for cubic, {\it etc}). This represents a
crucial difference between wave turbulence and hydrodynamic turbulence, because in the last one, the flux of energy appears in the second order correlation with the power $2/3$, which cannot be obtained as a $1/N$ power.

The turbulent asymptotic expansion of wave systems in the strongly nonlinear limit is, by construction, singular, because the linear term becomes then subdominant. However, a turbulent fully nonlinear dynamics is still expected to exist. The dependence of the statistical properties  with the potentially different nonlinear terms and the connexions with the NS fluid turbulence is then worth to be investigated. In this paper, we precisely address such singular limit in the case of vibrating elastic plates. This system has indeed recently revealed a strong analogy with hydrodynamic turbulence through the derivation of an exact law for a two-point correlation that is the equivalent of the Kolmogorov's 4/5-law~\cite{duringKrstulovic1Law}.

The weakly nonlinear turbulence regime in vibrating elastic plates has been established ten years ago by three of us~\cite{during}. Based on the dynamical F\"oppl--von K\'arm\'an equations 
\cite{foppl} (described below), we have shown using the WTT the existence of a spectrum of direct energy cascade.  The rather simple comparison between theory, numerics and experiments  has led to a large number of studies in vibrating plates (for a review see~\cite{platesPhysD}). Experiments performed soon after the theoretical predictions have shown slightly different spectrum power laws~\cite{arezki,mordant08}, that have been later explained by the specific features of the dissipation~\cite{EPL}.
Furthermore, wave turbulence of plates has shown to be a perfect system for investigating different concepts such as inverse cascade~\cite{inverse} and transitory dynamics~\cite{humbert16}.
More recently, thin plates with a high forcing~\cite{mordant13,japs}, the breakdown of the 
WTT and the onset of intermittency~\cite{Chibbaro} have been extensively studied. In these high
forcing regimes, a wave turbulence spectrum is still observed at small scales, while strong nonlinear regimes
appear at large scales, whose nature remain unclear, suggesting the existence of a strong turbulent dynamics.

In the following, to investigate this strongly nonlinear regime and the analogy with hydrodynamic turbulence, we will consider a thin elastic plate in the fully nonlinear case which corresponds to the formal limit with no bending and where linear waves are absent. With no surprise, the WTT spectrum vanishes in this limit and a different turbulent spectrum is expected to appear. The goal of the paper is to characterize such energy repartition and highlight similarities with hydrodynamic turbulence.

\section*{Theoretical model}

The vibration of a bending-free elastic plate comes from the usual dynamical version of the  F\"oppl--von K\'arm\'an equations 
\cite{landau} for the vertical amplitude of the deformation $\zeta(x,y,t)$ and for the Airy stress function $\chi(x,y,t)$ that describes the three in--plane stresses:
\begin{eqnarray} 
\rho\frac{\partial^2 \zeta}{\partial t^2} &=&  \zeta_{xx}\chi_{yy}+\zeta_{yy}\chi_{xx}-2\zeta_{xy}\chi_{xy} +\mathcal{F}+\mathcal{D};
\label{foppl0}\\
\frac{1}{E}\Delta^2\chi &=&- \left(  \zeta_{xx}\zeta_{yy} - \zeta_{xy}^2  \right).
\label{foppl1}
\end{eqnarray}
Here $\Delta=\partial_{xx}+\partial_{yy}$ is the usual Laplacian and $\rho$  and $E$ are, respectively, the mass density and the Young modulus $E$ of the plate. $\mathcal{F}$ and $\mathcal{D}$ are the forcing and dissipation respectively.

When $\mathcal{F}=\mathcal{D}=0$, the equations (\ref{foppl0}) and (\ref{foppl1}) derive from a Hamiltonian principle \cite{during,platesPhysD}. 
More important, it is a well posed system, in the sense that the energy is compound by two positive (hence bound from below) quantities, namely the kinetic energy and the stretching energy per unit mass:
 \begin{equation}
{ \cal E}_{\rm kin}= \frac{1}{2S} \int\dot \zeta^2 d{ \bf r}, \,\,{ \cal E}_{\rm stret}=  \frac{E}{8\rho S}\int\left[ \Delta^{-1} \left(  \zeta_{xx}\zeta_{yy} - \zeta_{xy}^2  \right) \right] ^2  d{\bf r},
\label{TwoEnergies}
\end{equation}
here  $S$ is the area plate.
For dimensional purposes, we use everywhere the energies per unit mass, ${ \cal E}$, ${ \cal E}_{\rm kin}$, and ${ \cal E}_{\rm stret}$, with dimensions of the square of a speed. 

\section*{Results} 
\subsection*{Turbulent behavior}
We performed numerical simulations of equations (\ref{foppl0}) and (\ref{foppl1}) by using a standard pseudo-spectral method in a periodic domain with an additive random forcing at large scales and a viscous damping acting at small scales. The explicit form of forcing and dissipation are discussed in Methods. Turbulent states should not depend explicitly on their details, provided that these mechanisms are well separated in the wavenumber space.

\begin{figure*}
\begin{centering}
\includegraphics[width=0.9\textwidth]{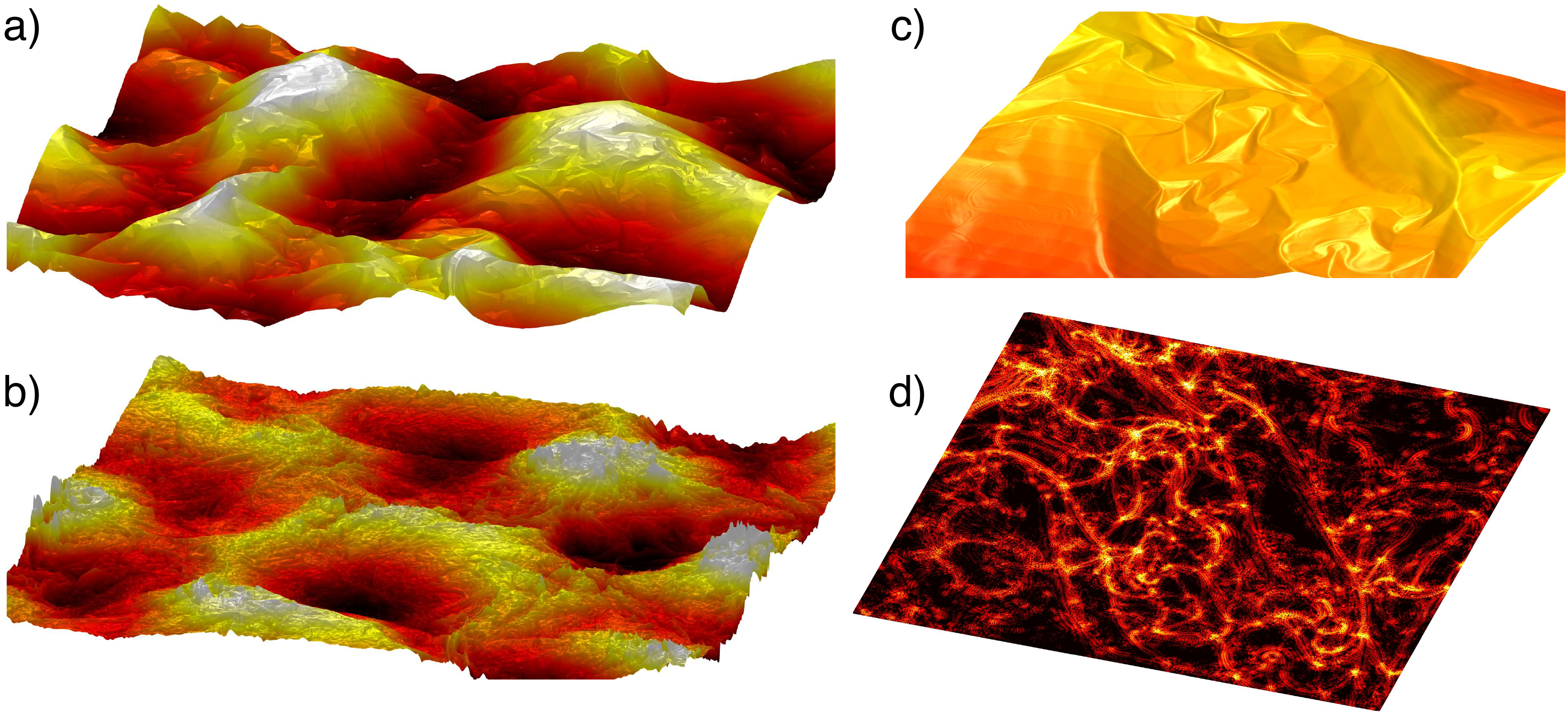}
\caption{Turbulent states of a bending-free plate. a) Surface plot of the plate deformation $\zeta(x,y)$.  b) Surface plot of the plate vertical velocity $\dot{\zeta}(x,y)$. c) Detail of the wrinkles in a zoom of the surface plot of the plate deformation $\zeta(x,y)$. d) Snapshot of the gaussian curvature \eqref{Gaussian} of the respective area. The density plot displays the values of $\log_{10} | G |$. Simulations were made at resolution of $4096^2$ with an additive random forcing, see Methods for more details.}\label{Fig:VizTurbulentstates}
\end{centering}
\end{figure*}

Figures \ref{Fig:VizTurbulentstates} a-b show the surface plate deformation $\zeta(x,y)$ and the plate vertical speed $\dot{\zeta}(x,y)$ respectively.
The displacement field displays a coarse scale, super-imposed with a fine scale fluctuations dominated by a large number of wrinkles of various size. Such multi-scale states are the most prominent characteristic of turbulent systems. The local speed of the plate in Fig.\ref{Fig:VizTurbulentstates}-b demonstrates that the elastic plate is not at equilibrium as it displays a myriad of excited modes. When bending is absent, it is natural to expect the appearance of highly non-linear geometrical structures. Such structures are visible in Fig. \ref{Fig:VizTurbulentstates}-c that shows a close-up of the plate deflection. 
The small scale wrinkles consist in a random assembly of moving ridges and conical points. Ridges and conical points are in fact the fundamental equilibrium configurations of elastic plates in the bending-free limit. 
These structures have been vastly studied since the pioneering works performed in the 90s \cite{Witten,Pomeau,DCone1}.  In the bending-free limit, the elastic deformations favor the bending modes because they cost no energy. Then, the deformations of the plates are controlled by the stretching that modifies the locally plane metric of the sheet. Because of geometrical constrains, it is not always possible to have a fully developable surface everywhere. The system thus creates singular structures: linear ridges \cite{Witten} and punctual developable cones or more commonly named D-cones \cite{Pomeau}, which concentrate the plate curvature and the stretching energy. Therefore, the dynamics of a bending-free elastic plate corresponds to a myriad of singularities moving (apparently) randomly over the sheet. In Supporting Information one can see a movie of these defect induced dynamics.   

To catch precisely these singular structures of dimension 1 (riges) and 2 (D-cones) we have plotted in Fig. \ref{Fig:VizTurbulentstates}-d the first order correction to the instantaneous Gaussian curvature:
 \begin{equation} G(x,y,t)= \zeta_{xx}\zeta_{yy} - \zeta_{xy}^2\label{Gaussian}
\end{equation}
 which, after \eqref{foppl1}, is the source of the in-plane stresses via the Airy function $\chi(x,y,t)$. The Gaussian curvature shows the complex network of ridges and D-cones.
 
\begin{figure}
\begin{center}
a) \includegraphics[width=0.2\textwidth]{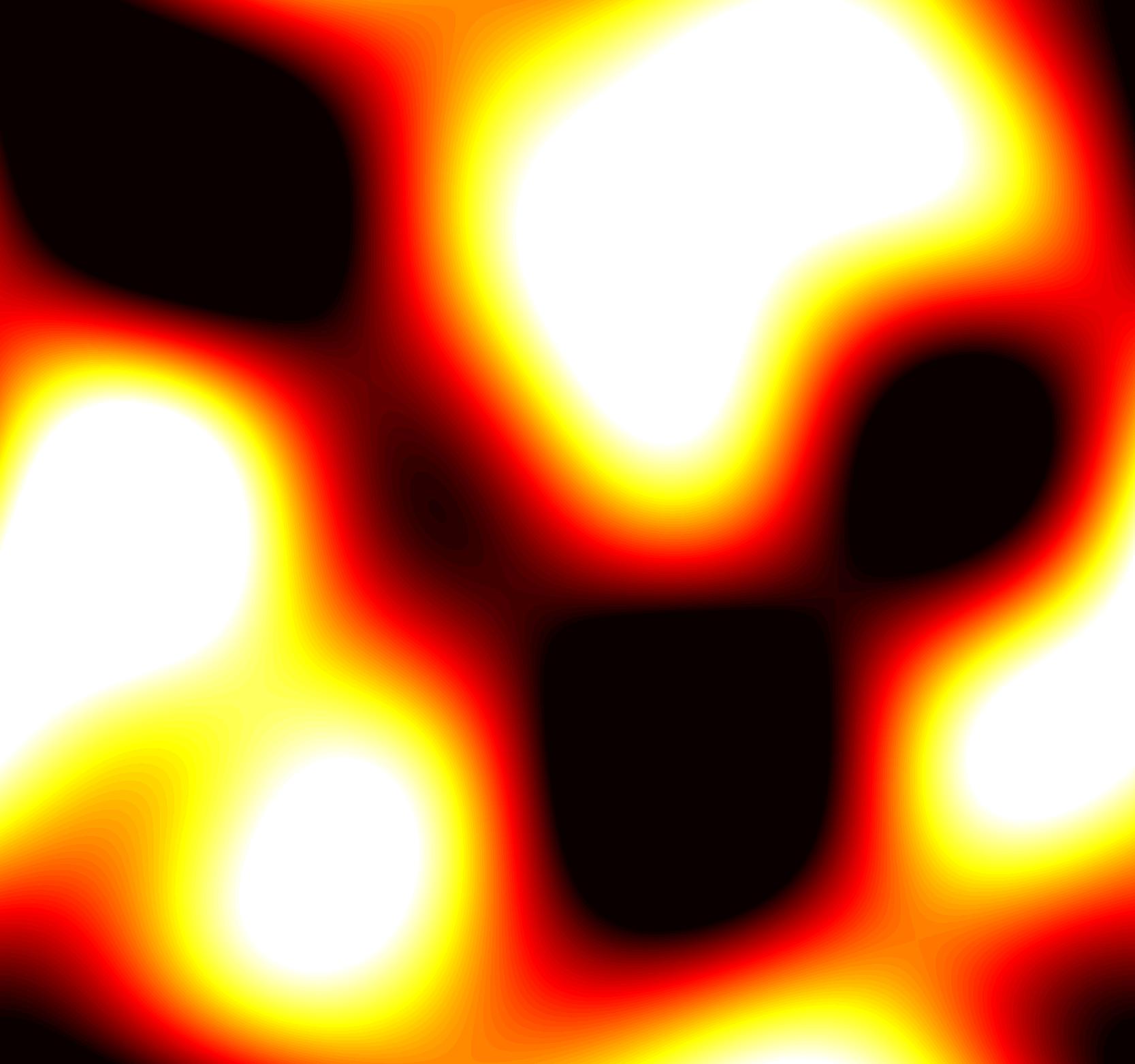}
b) \includegraphics[width=0.2\textwidth]{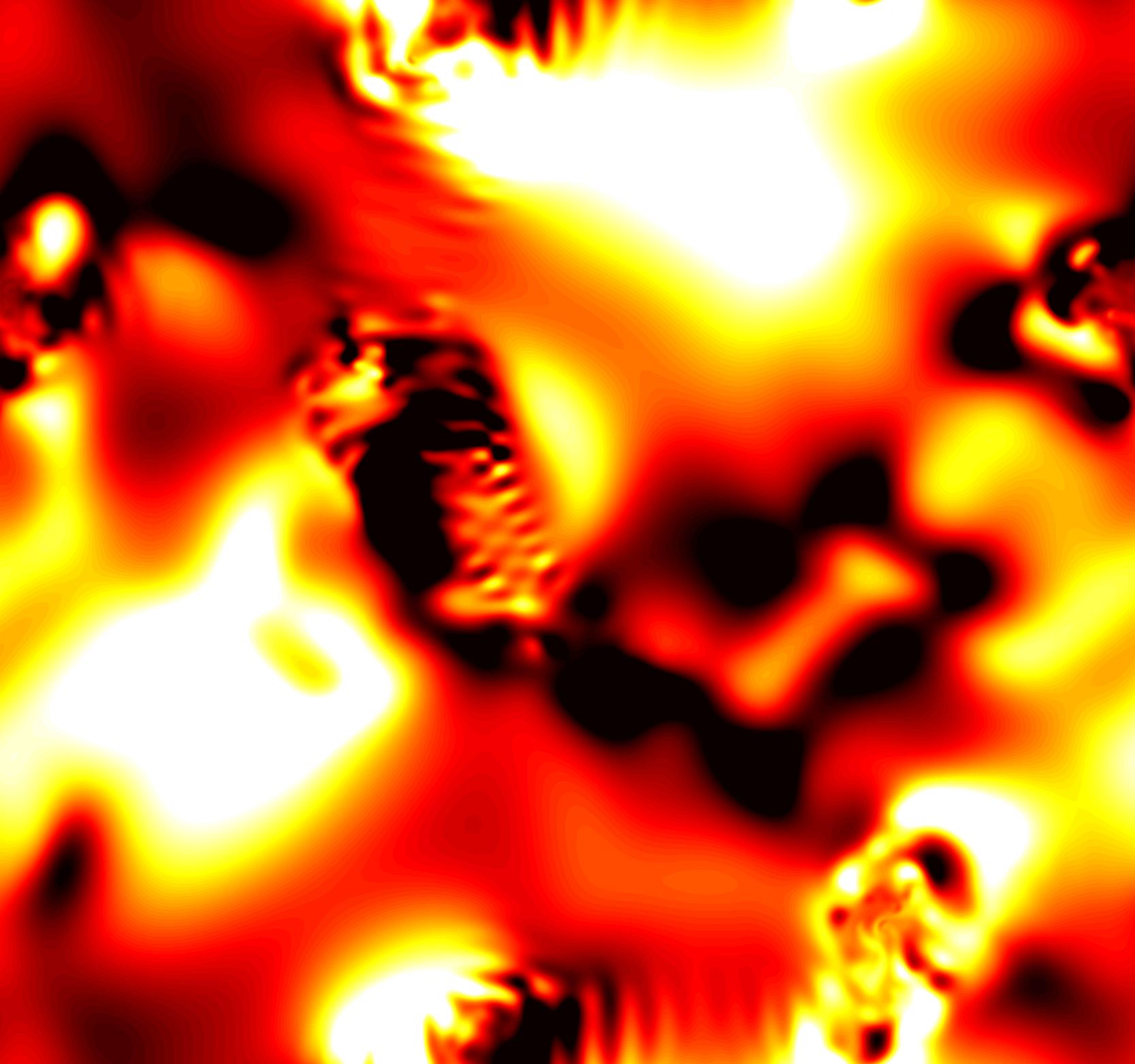}\\
c) \includegraphics[width=0.2\textwidth]{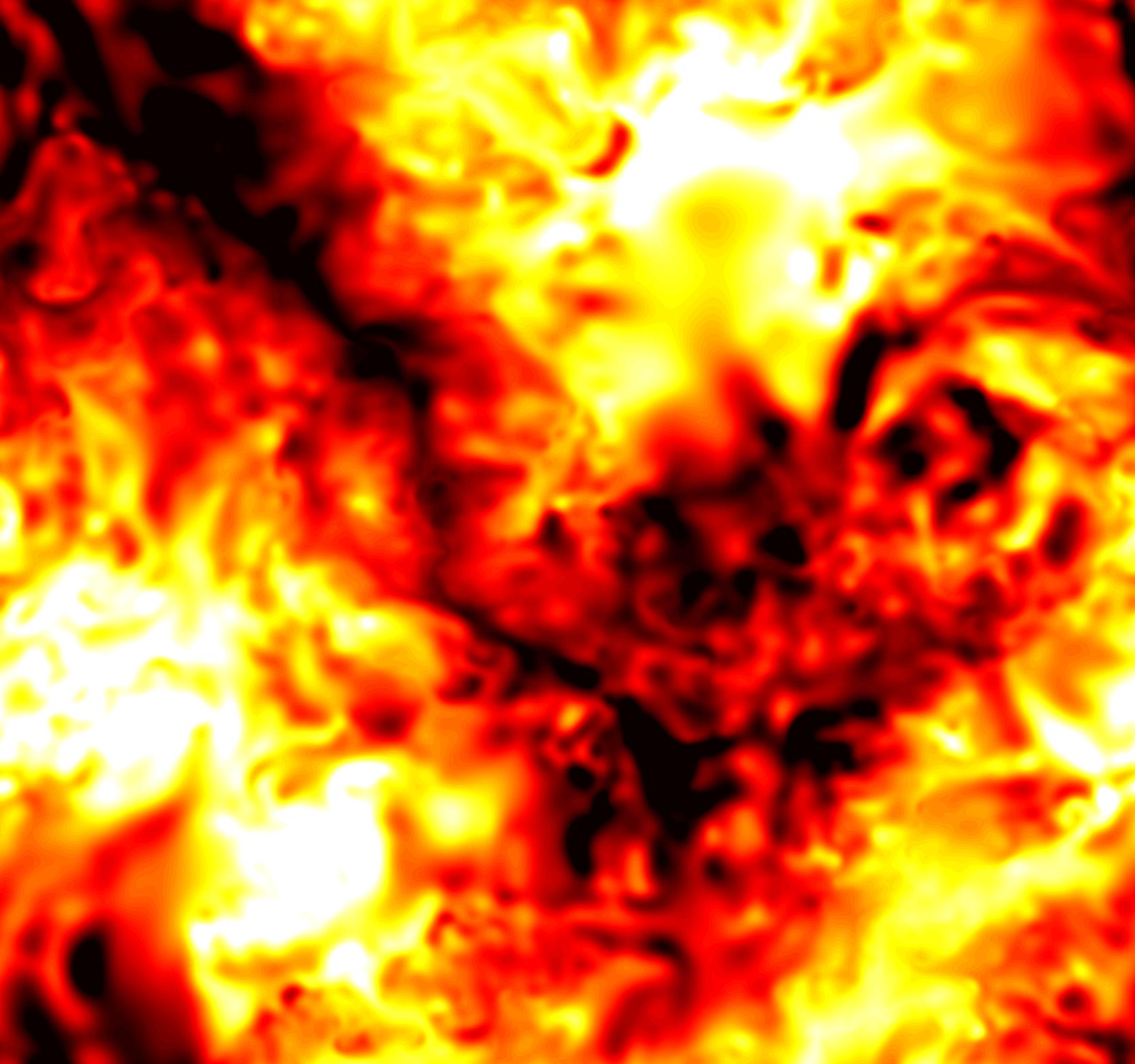}
d) \includegraphics[width=0.2\textwidth]{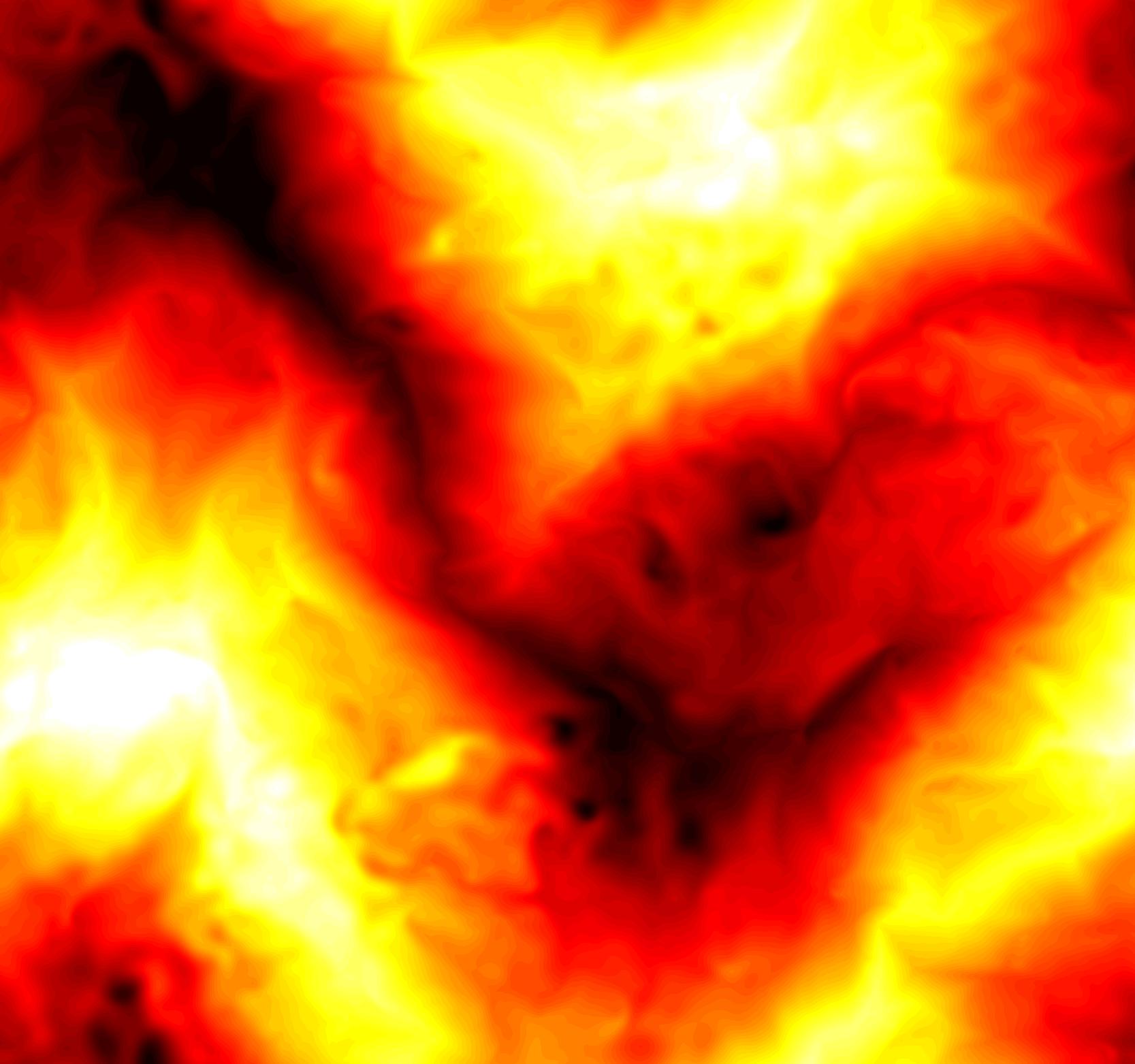}
\caption{Simulations for decaying turbulence. Snapshots of $\dot{\zeta}$. Time goes  as: a) $t=0$, b) $t=0.5$, c) $t=65,$ and d) $t=100$ (in time units  of \eqref{foppl0} and \eqref{foppl1}).}
\label{Fig:Decaying}
\end{center}
\end{figure}

An immediate analogy can be established with hydrodynamic turbulence. Fully nonlinear elastic plate turbulence seems to be characterized by a ``crumpling cascade'' where wrinkles play the role of whirls and instead of vortex filaments and sheets, the localized singularities come in the form of ridges and D-cones. To illustrate this idea further, we integrate the system \eqref{foppl0} and \eqref{foppl1} starting form a flat state, with no forcing  ($\mathcal{F}=0$), but with an initial velocity at large-scales and let the system decay. Snapshots of the velocity field, shown on Fig. \ref{Fig:Decaying}, exhibit the dynamics of this decaying turbulence configuration.
Initially, all the velocity fluctuations are contained at large scales Fig. \ref{Fig:Decaying}-a. Then, due to the non-linear mode interactions, instabilities appear creating smaller and smaller structures Fig. \ref{Fig:Decaying}-b. After this transient, a turbulent state is observed where fluctuations at all scales coexist Fig. \ref{Fig:Decaying}-c. Finally, at large times, the dissipative terms acting at small scales kill the turbulent state, smoothing out the fluctuations (see Fig. \ref{Fig:Decaying}-d).

\subsection*{Kolmogorov spectrum} 

\begin{figure}
\includegraphics[width=0.45\textwidth]{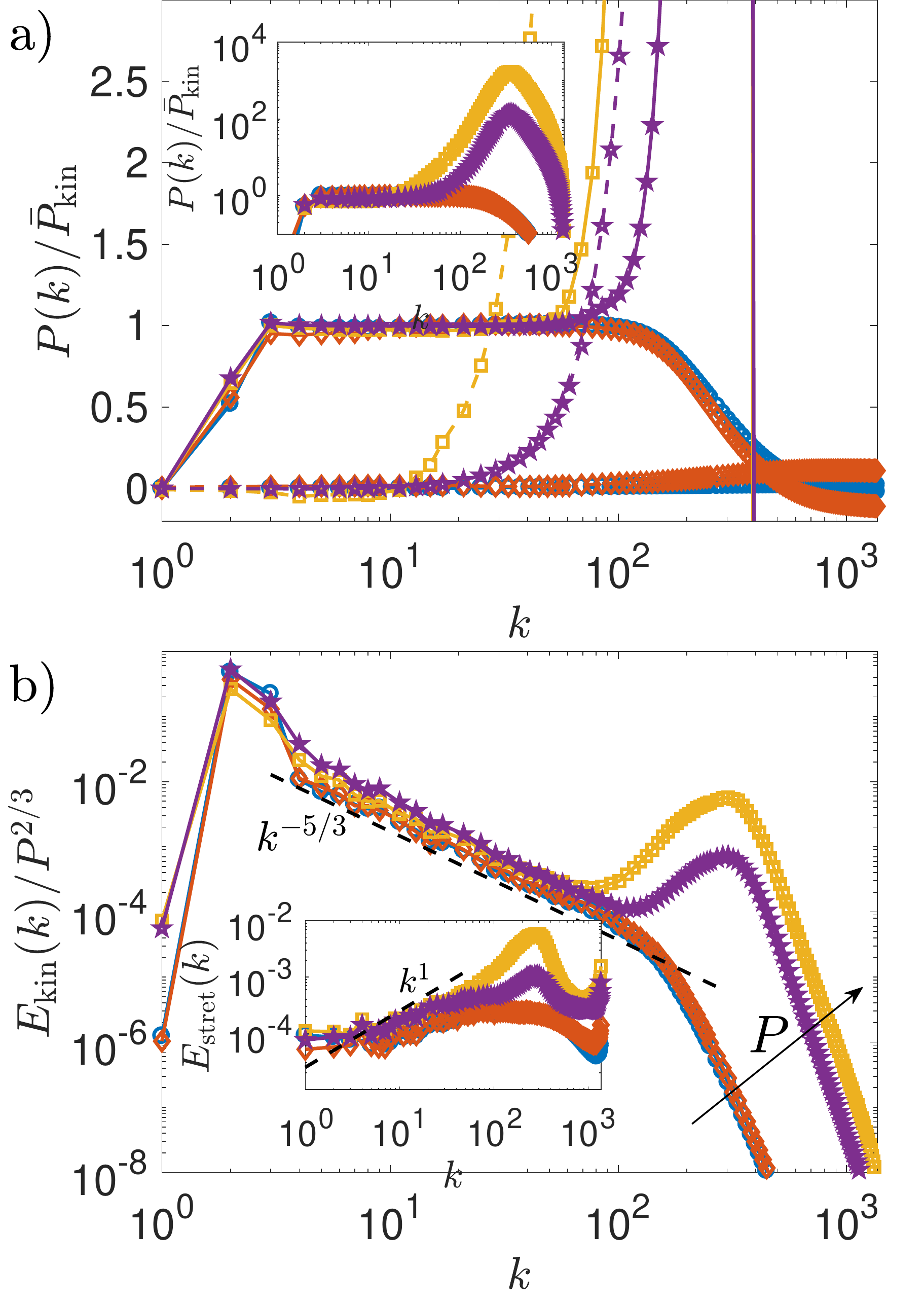}
\caption{
 a) Time averaged kinetic energy flux (solid lines) and the stretching energy flux (dashed lines). Different markers (and colors) are for different runs with increasing forcing amplitude or fluxes (see arrow in figure b). The fluxes have been normalized by their total mean flux value in the inertial range (see methods for values). The inset displays the same plot but in log-log.  b) Time averaged kinetic energy spectra normalized by $P^{2/3}$. The dashed line displays the Kolmogorov scaling. Inset: Time averaged stretching energy. The dashed line displays the thermalization energy scaling. The arrow indicates the different runs with increasing values of the flux $P$. The value of $\bar{P}_{\rm kin}$ varies in the range $ (0.03,16.5)$ for all the runs (see Methods). } 
\label{Fig:ForcedSpectraFlux}
\end{figure}

By forcing at large scales and dissipating at small scales, a turbulent out-of-equilibrium steady state is obtained, such as the one observed in Fig. \ref{Fig:VizTurbulentstates}. A quantitative measurement of these steady turbulent states is obtained through the spectral densities of the plate velocity. As in fluid turbulence, here we compute the average kinetic energy spectrum and the kinetic energy flux for different wave numbers. The kinetic energy spectrum per unit mass, $E_{\rm kin}(k)$, is defined through the kinetic energy per unit mass by
$$ {\cal E}_{\rm kin} = \int E_{\rm kin}(k) \,dk, $$
  where $E_{\rm kin}(k) =2 \pi k \left< |\dot{\zeta}_k|^2 \right>$, the velocity field  $\dot{\zeta}_{ k} (t) = \int \dot{\zeta}({\bf r},t) e^{i { \bf k}\cdot {\bf r}} d{\bf r}$ and isotropy is assumed. In addition, the kinetic energy flux $P_{\rm kin}(k)$ is defined through the transfer equation:
 \begin{equation} 
 \frac{\partial}{\partial t}E_{\rm kin}(k) = -\frac{\partial}{\partial k} P_{\rm kin}(k).
 \label{EnergyFluxDef}
  \end{equation} 
Since  $E_{\rm kin}(k)$  has dimensions of $ L^{3}/T^2$  the dimensions of $P_{\rm kin}(k)$ is $L^2/T^3$. Notice that the stretching energy \eqref{TwoEnergies} is a quadratic quantity, therefore analogous definitions can be given for its spectrum and flux (See Methods for explicit definitions). 
We have measured directly the energy fluxes for various turbulent runs with different forcing amplitudes. Figure \ref{Fig:ForcedSpectraFlux}-a shows the time-averaged kinetic (solid lines) energy fluxes, normalized by their mean value $\bar{P}_{\rm kin}$ in the well defined transparency window where they are flat. 
This transparency window is usually called the inertial-range in the context of hydrodynamic turbulence. At small scales, dissipation takes place and produces a bottleneck that invades the inertial range as the mean energy flux increases, although an inertial range is still clearly present. This bottleneck is related to dissipative effects, that have been shown to be non-trivial \cite{EPL}. Figure~\ref{Fig:ForcedSpectraFlux}-a also displays the stretching energy fluxes (dashed lines), which are notably smaller than the kinetic energy fluxes. Therefore, numerical simulations indicate that the pertinent energy flux for the cascade is the kinetic energy one, because it shows a constant energy flux along scales in an inertial range. A constant energy flux  is usually associated with a turbulent regime that leads to a power law behavior for the corresponding energy spectrum. This is indeed the case for the kinetic energy spectrum, as shown in Fig. \ref{Fig:ForcedSpectraFlux}-b where the $k^{-5/3}$ Kolmogorov law is apparent. 

The kinetic turbulent spectrum can be discussed on dimensional grounds in the footprint of Kolmogorov. However, here in addition to the constant energy flux per unit mass 
$P$ and the wavenumber  $k$ (with dimension inverse of a length), the spectrum should depend {\it a priori} 
on the
additional material parameter $E/\rho$. Since a characteristic length can be defined, namely $\lambda = (E/\rho)^{3/2} /{P}$, one gets generically 
\begin{equation}
E_{\rm kin}(k)= \frac{P^{2/3}}{ k^{5/3} } \Phi\left( k  \frac{(E/\rho)^{3/2} }{P}  \right),
\label{spech0}
\end{equation}
where $\Phi(\cdot)$ is an arbitrary function of the dimensionless argument $ k\lambda$.
Contrary to fluid turbulence, the existence of this extra length $\lambda$, does not allow to uniquely determine a fixed exponent for the power law spectrum. Remarkably, numerics suggest that the kinetic and the stretching energy spectra (\ref{Methods.Spectra}) can be treated independently. Considering that for the kinetic energy  the elastic property of the material does not intervene, the kinetic energy spectrum should follow
\begin{equation}
E_{kin}(k)=C_{kin} P^{2/3} k^{-5/3} , 
\label{spech1}
\end{equation}
with $C_{kin} = \Phi(0)\neq 0$ a constant. The numerical simulations show indeed a good agreement with this predicted scalings \eqref{spech1} for both $k$ and $P$.

Because the energy \eqref{TwoEnergies} combines two different contributions we address the role of the stretching energy.  We show (See methods for the proof), via a simple argument  that the stretching energy flux related to the stretching energy must vanish in a statistically steady state under an additive forcing and a viscous-like dissipation, as the one considered in this work.  This result follows directly from the observation that the stretching energy flux appears to be the time derivative of the stress correlation function which must be zero in steady state turbulent regime. Therefore, one would expect the stretching modes to eventually thermalize. Since the stretching energy is quadratic in $\gamma=\Delta^{-1} \left(  \zeta_{xx}\zeta_{yy} - \zeta_{xy}^2 \right)$ (see \eqref{TwoEnergies}), the equilibrium distribution corresponds to equipartition of the Fourier modes of $\gamma$, leading to the equipartition spectrum of the form:
\begin{equation}
E_{stret}(k)= C_{stret} k , 
\label{spech2}
\end{equation}
with $C_{stret}$ a constant proportional to the mean energy, that plays the role of an effective temperature. In the inset of Fig. \ref{Fig:ForcedSpectraFlux} the stretching spectra for different forcing amplitude are shown to be consistent with the equipartition law \ref{spech2}, for the scales where the dissipation is negligible.

\subsection*{Intermittency and beyond Kolmogorov phenomenology}

The Kolmogorov phenomenology discussed in the previous section is based on dimensional analysis and mean-field assumptions that neglect the existence of extreme fluctuations. Nevertheless, it is well known that deviations exist to such predictions and they become important when looking at higher order statistics~\cite{FrischBook}. These deviations are associated with the intermittent statistics of the fields and are 
somehow inherent to fluid turbulence. A complete understanding of such fluctuations is still missing. Unlike hydrodynamic turbulence, the theory of weak wave-turbulence predicts Gaussian statistics for the distribution of wave amplitudes, therefore no intermittency can be observed within the range of validity of the theory. However, when the non-linearities become of the same order than the linear dispersive terms, the wave turbulence theory breaks down and an intermittent statistics can manifest~\cite{NewellNazarenkoIntermittency}. Such intermittencies have been experimentally observed in gravity-capillary waves~\cite{FalconIntermittency2007} and suggested for thin elastic plates~\cite{MordantIntermittencyPlate}. Moreover, numerical simulations of elastic plates at strong forcing have shown important deviations from the WTT predicted scalings for the plate deformation~\cite{japs,Chibbaro}, leading to intermittency signatures~\cite{MordantIntermittencyPlate,Chibbaro}.

Intermittency is usually addressed looking at the moments of the so-called structure functions of the fields, that provide information of the variation of the fields at a given scale \cite{FrischBook}. For elastic plates, because of the fast decay of the deformation spectrum, a second order difference is needed to observe an intermittent scaling~\cite{Chibbaro}. In our system, where bending waves are absent and the dynamics is thus fully non-linear, the intermittent behavior is expected to occur at any forcing. Following~\cite{Chibbaro}, we introduce the (second variation) increments $\delta^2_{ \ell} \zeta$ of the plate deformation and their corresponding structure functions $S_p (\ell)$, namely
\begin{eqnarray}
S_p (\ell) = \langle | \delta^2_\ell \zeta|^p\rangle \,\, {\rm with} \,\,
\delta^2_{ \ell} \zeta =\zeta({\bf x} + {\bf \ell})-2\zeta({\bf x})+ \zeta({\bf x} - {\bf \ell}).
\label{def:Sp}
\end{eqnarray}

\begin{figure}
\includegraphics[width=0.45\textwidth]{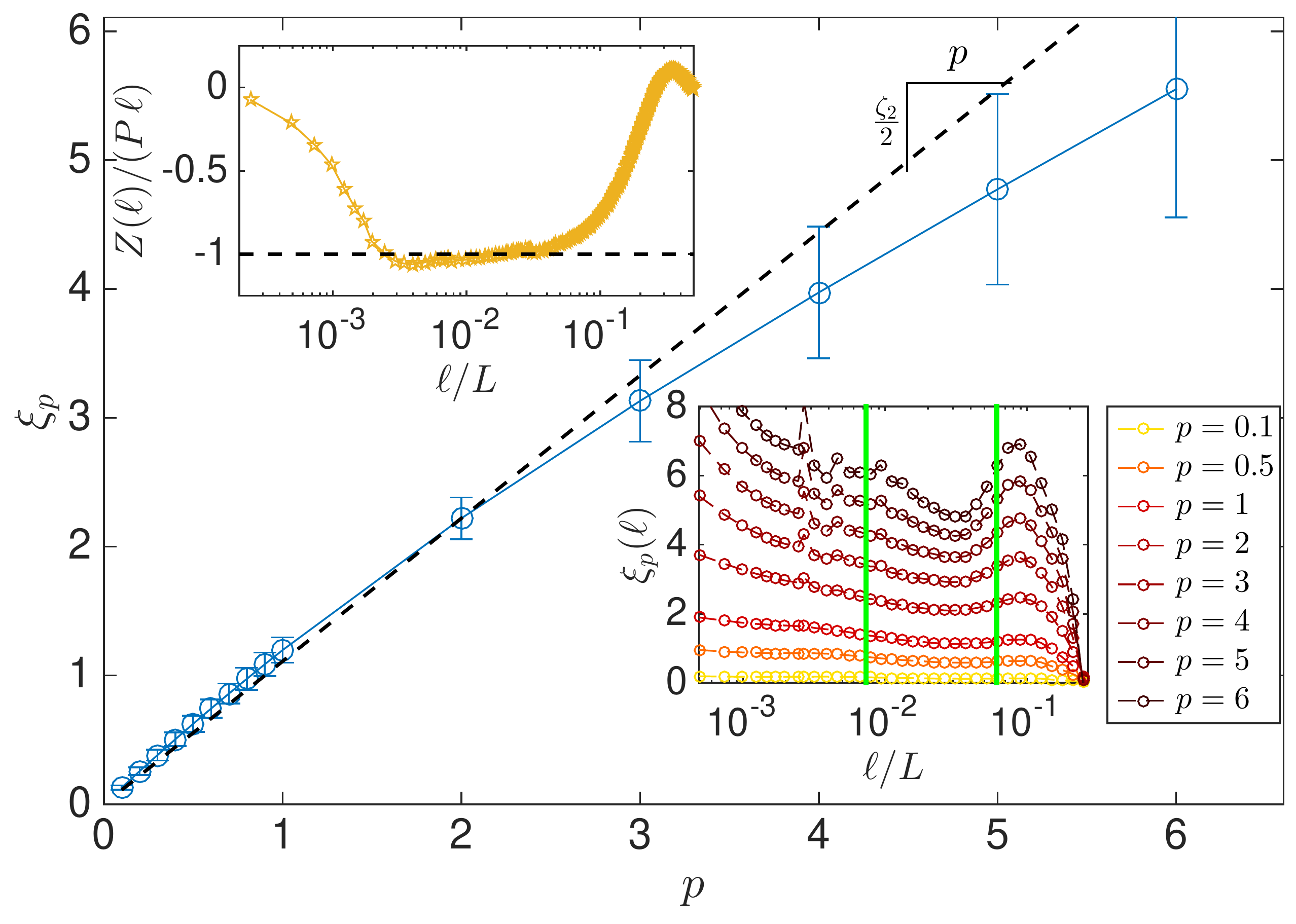}
\caption{Intermittency exponents $\xi_p$ as a function of $p$ showing clearly a departure from a linear law, indicating the existence of anomalous exponents and intermittency. The right down inset shows the structure functions local slopes as a function of $\ell$. The inertial range is delimited by the two vertical green lines. The left-up panel shows evidence of the 1-law \eqref{Eq:1law}. The numerics was done under the same conditions as previous figures. }
\label{Fig:Intermittency}
\end{figure}

At very small scales, the regularity of $\zeta({ x})$ implies the scaling  $S_p (\ell)\sim\ell^{2 p}$. However, in the inertial range a non-trivial scaling can appear. We define then the anomalous exponents (in the spirit of hydrodynamic turbulence) $\xi_p$, from the moments following
$
S_p (\ell)\sim \ell^{\xi_p}.
$
From our numerics we measure that $\xi_2\approx 2.2$. In Kolmogorov theory as well as for WWT, a linear relation between the different exponent is expected $\xi_p=\frac{\xi_2}{2}p$, witnessing the self-similar nature of the dynamics. For vibrating elastic plates however, a deviation from this linear law has been observed, 
indicating thus clearly the presence of intermittency that breaks the self-similarity of  crumpling dynamics~\cite{Chibbaro}. In the bottom inset of Fig. \ref{Fig:Intermittency}, we present the local slope of the structure functions defined as $\xi_p(\ell)=\frac{d \log{S_p (\ell)}}{d \log{\ell}}$ for different orders $p$. 
A relatively flat behavior is observed in the inertial range and the anomalous exponents $\xi_p$ are measured by averaging $\xi_p(\ell)$ in this window.  Fig. \ref{Fig:Intermittency} shows $\xi_p$ as a function of $p$, exhibiting a clear departure from the linear behavior (dashed line) and thus intermittent statistics.  
Moreover,  we have verified the exact result derived by two of us for thin elastic plates that is valid for both, weak and strong wave turbulence \cite{duringKrstulovic1Law}. This result is the equivalent to the only exact result of hydrodynamic turbulence known as the $4/5$-Kolmogorov law of turbulence. It relates increments of the field with the energy flux and the scales. In the case of a thin elastic plate, a similar relation was found for a different structure function, called $Z(\ell)$, that depends on the first order variation of the fields $\chi$, $\zeta$ and $\dot{\zeta}$ (see Methods). In analogy with the $4/5$-Kolmogorov law, this relation was called the $1$-law and reads
\begin{equation}
Z(\ell)=-P \ell. \label{Eq:1law}
\end{equation}
The structure function $Z(\ell)$, normalized by $P\ell$ is displayed in (top) inset of Fig. \ref{Fig:Intermittency}. An excellent agreement with the prediction \eqref{Eq:1law} is observed over a decade with no adjustable parameter.
Finally, we have also studied the vertical speed structure function $\left< \left(\dot\zeta({\bf r}+{\bf \ell})- \dot\zeta({\bf r} ) \right)^3\right>$. Unlike, hydrodynamic turbulence, it does not scales linearly with the distance $\ell$. This difference can be explained by the existence of the extra-length $\lambda$, suggesting that the exponent can not be uniquely determined by dimensional analysis. This fact is also related to the complexity of the non-linear term, that is precisely taken into account in \eqref{Eq:1law}.

\section*{Discussion}\label{sec:conclusion}

Substantial evidence shows robustness among turbulent behavior in ordinary fluids and in the case of crumpling vibrations of an elastic plate in the zero limit thickness. As presented, both manifest an energy cascade with the well known K41 spectrum $E_k\sim P^{2/3} k^{-5/3}$. More important, a myriad of randomly interacting highly nonlinear crumpling structures (folds, ridges and d-cones) at all relevant scales dominate the dynamics. They induce intermittency that we have quantitatively investigated by studying high-order correlators that confirm the appearance of an intermittent behavior. 
Nevertheless, the underlying plate dynamics differs notably from the one of ordinary fluids: though in incompressible fluids the total energy consists purely of kinetic energy, in elastic plates the energy is compound of two separately (positive) energies: the kinetic and the stretching energy. Therefore, it is expected that two different cascades may exist in the elastic situation, a kinetic energy one and a stretching energy one. 
Because the forcing is additive to the deformation acceleration \eqref{foppl0}, only a constant kinetic energy flux (see  Fig. \ref{Fig:ForcedSpectraFlux}-a) exists resulting in a K41 cascade. On the other hand, the stretching energy flux vanishes (see  Fig. \ref{Fig:ForcedSpectraFlux}-a) and consequently Gibbs equipartition is observed for the stretching spectrum, as shown in the inset of  Fig. \ref{Fig:ForcedSpectraFlux}-b.  Naturally it is expected that other forcing may display a stretching energy cascade without any kinetic energy cascade. Or perhaps  two distinct cascades a kinetic energy K41 cascade and a stretching cascade simultaneously. 

\appendix

\section*{Appendix}
{
In the following we describe the mathematical and numerical methods used in the present paper.
\subsection*{Numerics}
We solve the  F\"oppl--von K\'arm\'an equations \eqref{foppl0}-\eqref{foppl1} with a standard pseudo-spectral code in a square domain of size $2\pi$ with periodic boundary conditions.
The forcing $\mathcal{F}$  is white-noise in time of variance $f_0^2$ and its Fourier modes are non-zero only for wave-vectors $1.5\le|{ k}|\le 3$. De-aliasing is made by using the standard $2/3$-rule, that is applied after computing each quadratic term. The dissipation is acting at  small and large scales in the following way
\begin{eqnarray}
\mathcal{D} &=&- \nu(-\Delta)^{n_\nu}\dot{\zeta} - \alpha(-\Delta)^{-n_\alpha}\dot{\zeta}
\label{num:addnoise}
\end{eqnarray}
The run presented in Fig. \ref{Fig:Decaying} was performed with  $1024^2$ collocation points, $\mathcal{F}=0$, ${n_\nu}=2$, $\nu=4\times 10^{-6}$, $\alpha=0$ and a random initial condition at large scale only on $\dot{\zeta}$. All the other runs presented in this work were performed with $4096^2$ collocation points, ${n_\nu}=3$,  $\nu=2\times 10^{-12}$, $n_\alpha=2$, $\alpha=100$ and $f_0=1, 8, 27$ and $100$. For the different forcing amplitudes, the measured value of the fluxes were $0.0284,0.0354,5.3358$ and $16.5325$ respectively.

\subsection*{The energy spectra}
The energies in \eqref{TwoEnergies} maybe seen as quadratic contribution of $\dot\zeta$ and  $ \gamma({\bf r}) = - \Delta^{-1} \left(   \zeta_{xx}\zeta_{yy} -  \zeta_{xy}^2 \right)$.The final quadratic contributions read
\begin{eqnarray}
{ \cal E}_{\rm kin}= \frac{1}{2} \int | \dot \zeta_{\bf k} |^2 d{\bf k} \quad {\rm and} \quad { \cal E}_{\rm stret}=  \frac{E}{2\rho} \int |  \gamma_{\bf k} |^2 d{\bf k}. \label{Methods.Energies}
\end{eqnarray}
We  define the kinetic energy and the stretching spectra by averaging over the angular variables  in the wavenumber space, hence
\begin{eqnarray}
E_{\rm kin}(k) = \pi k  \left< | \dot \zeta_{ \bf k} |^2  \right> \quad {\rm and}\quad 
 { E}_{\rm stret} (k)= \pi \frac{E}{\rho}  k   \left<|  \gamma_{ \bf k} |^2 \right>,\label{Methods.Spectra}
\end{eqnarray}
where we have defined  $\left<\dots \right>= \frac{1}{2\pi} \int \dots d\varphi_k$ by the angular average.

\subsection*{The energy fluxes}
Because the energy is quadratic in $\dot{\zeta}$ and $\gamma$, the energy flux can be straightforward defined as in hydrodynamic turbulence. By making a scale-by-scale energy budget  the energy fluxes are: 
 \begin{eqnarray}
P_{\rm kin}(k)=-\int_0^k \left.\frac{\partial E_{\rm kin}(p)}{\partial t}\right|_{0}dp=-\int_0^k  \left<\dot{\zeta}_{ \bf p}\{\zeta,{ \chi}\}_{- \bf p} \right> dp ,\label{KineticEnergyFlux} \\
P_{\rm stret}(k)=-\int_0^k \left.\frac{\partial E_{\rm stret}(p)}{\partial t}\right|_{\rm 0}dp =\int_0^k \left<\chi_{\bf p} \{\zeta,\dot{\zeta}\}_{- \bf p}\right> dp,\label{StretchingEnergyFlux}
\end{eqnarray}
where the subscript 0 stands for the free of forcing and dissipation time variations of the fields through \eqref{foppl0} with ${\mathcal F}={\mathcal D}=0$. 
For simplicity, we define $\{f,g\}\equiv f_{xx}g_{yy}+f_{yy}g_{xx}-2f_{xy}g_{xy}$. We have also used the fact  that the angular averages satisfy $\left< (\Delta^{-1} f)_{ \bf p} (\Delta^{-1} g)_{- \bf p} \right> =\left< (\Delta^{-2} f)_{ \bf p}  g_{- \bf p} \right>$. 
The total energy flux results from the addition of these previous expressions.

\subsection*{Steady state with zero stretching energy flux}
Following \cite{duringKrstulovic1Law}, we introduce the spatial correlation function
 \begin{eqnarray}
\label{stretching}
\mathcal{E}_\text{stret}({ \bf \ell})= \frac{1}{2E}\langle\Delta_{ r}\chi({\bf r})\Delta_{{\bf r}'}\chi({ \bf r}')\rangle ,
\end{eqnarray}
with ${ \ell}={\bf r}'-{ \bf r}$, which on the limit of $\ell \rightarrow 0 $ converges to the mean value of the stretching energy. Considering an homogeneous and isotropic system, then the time derivative of the correlation function \eqref{stretching} can be shown to be 
\begin{eqnarray}
\mathcal{\dot{E}}_\text{stret}({ \ell})= -\langle\chi({\bf r}')  \{\zeta({\bf r}),\dot{\zeta}({ \bf r})\} \rangle 
=\frac{1}{V}\int\chi({\bf r}+{\bf \ell})  \{\zeta({ \bf r}),\dot{\zeta}({ \bf r})\}d{ \bf r}.
\end{eqnarray}
 The last equality considers that for an homogeneous system the statistical average can be taken as a spatial average. From the convolution theorem the Fourier transform ${\mathcal{\dot{E}}_\text{stret}}({\bf p})=\chi_{ \bf p}\{\zeta,\dot{\zeta}\}_{- \bf p}$. In a steady state the time derivative of the correlation function must be zero hence $\chi_{\bf p}\{\zeta,\dot{\zeta}\}_{-\bf p}=0$ and the stretching energy flux \eqref{StretchingEnergyFlux} must be zero.
\subsection*{1-law for thin elastic plate turbulence}
In Ref. \cite{duringKrstulovic1Law} an exact result for turbulence of thin elastic plates was found. It provides a  K\'arm\'an-Howarth-Monin type relationship for the energy flux:
\begin{equation}
\frac{1}{2}\nabla_{\bf \ell}\cdot  \langle  {\bf J}_{[\delta\chi,\delta\zeta]}\delta\dot{\zeta}\rangle=-P\label{KHM_Intertial},
\end{equation}
where $\delta$ stands for the first difference of the field, {\it e.g.} $\delta\zeta=\zeta({ \bf x}+{ \bf \ell})-\zeta({ \bf x})$. The vector ${ \bf J}$ is defined by  
$$
{ \bf J}_{[ f ,h]}= f_y h_{yx}-f_{x}h_{yy} , f_x h_{xy} - f_{y} h_{xx}.
$$
Under the assumption of isotropy, \eqref{KHM_Intertial} implies the \emph{$1$-law} \eqref{Eq:1law} for the structure function 
\begin{equation}
Z(\ell)\equiv \langle {\bf J}_{[\delta\chi,\delta\zeta]}\delta\dot{\zeta}\rangle\cdot \hat{{\bf \ell}}=-P\,  \ell, \label{EQ:OneLaw}
\end{equation}
where $\hat{\bf\ell}$ is the unit vector along ${\bf \ell}$.

}

\begin{acknowledgements}
GD, GK and SR thank to FONDECYT grants N$^\circ$ 1181382 and  the Chilean-French scientific exchange program ECOS-Sud/CONICYT No. C14E04.
\end{acknowledgements}

\bibliographystyle{}

\end{document}